\documentclass[preprint,preprintnumbers,superscriptaddress,amsmath]{revtex4}

\usepackage[dvips]{graphics}
\usepackage{epsfig}
\usepackage{psfrag}
\usepackage[psamsfonts]{amssymb}
\usepackage{amsmath}
\usepackage{indentfirst}
\usepackage{amssymb}
\usepackage{wrapfig}
\usepackage{dcolumn}
\usepackage{bm}

\newcommand{\be}{\begin{equation}}
\newcommand{\ee}{\end{equation}}
\newcommand{\ba}{\begin{eqnarray}}
\newcommand{\ea}{\end{eqnarray}}

\begin{document}

\preprint{APS preprint}

\title{Renormalization of the ETAS branching model of triggered seismicity
from total to observable seismicity}
\author{A. Saichev}
\affiliation{Mathematical Department, Nizhny Novgorod
State University, Gagarin prosp. 23, Nizhny Novgorod,
603950, Russia} \affiliation{Institute of Geophysics and
Planetary Physics, University of California, Los Angeles,
CA 90095}

\author{D. Sornette}
\affiliation{Institute of Geophysics and Planetary
Physics and Department of Earth and Space Sciences,
University of California, Los Angeles, CA 90095}
\affiliation{Laboratoire de Physique de la Mati\`ere
Condens\'ee, CNRS UMR 6622 and Universit\'e de
Nice-Sophia Antipolis, 06108 Nice Cedex 2, France}
\email{sornette@moho.ess.ucla.edu}

\date{\today}

\begin{abstract}

Several recent works point out that the crowd of small unobservable
earthquakes (with magnitudes below the
detection threshold $m_d$) may play a significant and perhaps dominant role
in triggering future seismicity. 
Using the ETAS branching model of triggered seismicity,
we apply the formalism of generating probability functions
to investigate how the statistical properties of observable
earthquakes differ from the statistics of all events. 
The ETAS (epidemic-type aftershock sequence) model
assumes that each earthquake can trigger other earthquakes
(``aftershocks''). An aftershock sequence results in this model from the
cascade of aftershocks of each past earthquake. The triggering
efficiency of earthquakes is assumed to vanish below 
a lower magnitude limit $m_0$, in order to ensure the convergence
of the theory and may reflect the physics of state-and-velocity frictional
rupture. We show that, to a good approximation, the ETAS model
is renormalized onto itself under what amounts to 
a decimation procedure $m_0 \to m_d$, with just a renormalization of the 
branching ratio from $n$ to an effective value $n(m_d)$. 
Our present analysis thus confirms,
for the full statistical properties, the results obtained previously
by one of us and Werner, based
solely on the average seismic rates (the first-order moment of the
statistics). However, our analysis also demonstrates that this
renormalization is not exact, as there are small corrections
which can be systematically calculated, in terms of additional
contributions that can be mapped onto a different branching model
(a new relevant direction in the language of the renormalization group).

\end{abstract}

\pacs{64.60.Ak; 02.50.Ey; 91.30.Dk}

\maketitle

\section{Introduction}

In the last few years, physicists' interest
for the  space-time organization of seismicity in different
regions of the world has spurred. This recent burst of attention is
probably due to the introduction of new
diagnostic tools applied to earthquake catalogs
\cite{BaketalOmo,Lindman_descaled,ergotiampo,Mega,helmsorcomment,network,
Corral,Corral2,Corral3,paczuski,paczuski2,paczuski3,paczuski4,
HSdiff2,MarsanBean,MarsanBean2,scaffetawest,YangSOC,YangSOC2} and to improved
insights from cartoon models of earthquakes
\cite{rundlefils,Herneu,HSdiff1,herhelmsor,SO1_mf,OS1_mf}. The first
class of papers in particular suggest to re-examine the standard statistical
properties of earthquakes, usually documented under
the following distinct power law and fractal properties:
(i) the Gutenberg-Richter distribution
$\sim 1/E^{1+\beta}$ (with $\beta \approx 2/3$) of
earthquake energies $E$ \cite{KKK}; (ii) the Omori law $\sim 1/t^p$ (with
$p \approx 1$ for large earthquakes) of the rate of
aftershocks as a function of time $t$ since a mainshock \cite{utsu};
(iii) the productivity law $\sim E^{a}$ (with $a \approx 2/3$)
giving the number of earthquakes triggered by an event of
energy $E$ \cite{H}; (iv) the power law distribution  $\sim
1/L^2$ of fault lengths $L$ \cite{Davy}; (v) the fractal
structure of fault networks \cite{davy2} and of the
spatial organization of earthquake epicenters \cite{KK}; (vi)
the distribution $1/s^{2+\delta}$ (with $\delta \geq 0$)
of seismic stress sources $s$ in earthquake focal zones
due to past earthquakes \cite{kagan94}. 
Specifically, the statistical analysis based on (a)
coarse-grained scaling ansatz \cite{BaketalOmo,Lindman_descaled,
Corral,Corral2,Corral3,paczuski4}
(b) entropic methods \cite{Mega,MarsanBean}, and (c) network methods 
\cite{network,paczuski,paczuski2,paczuski3}
suggest that the above standard seismological description 
\cite{KKK,utsu,H,Davy,davy2,KK,kagan94} may be inadequate. It is not
clear however what should be the correct physical model. 
Several papers have however questioned the novelty of the insights derived
from these approaches \cite{Lindman_descaled,helmsorcomment,MarsanBean2}.

The present authors are among those who have studied
how the standard seismological laws \cite{KKK,utsu,H,Davy,davy2,KK,kagan94} 
(in particular the laws (i)-(iii) mentioned above)
could actually go a long way towards explaining most of the empirical phenomenology
of seismicity, including the supposed anomalous or ``novel'' scaling 
laws proposed by the above quoted physicists (see for instance
\cite{HS03,HSbath,Saichevetal04,SaichSorl04,SSbath,SaiSorpdf}).
In this series of papers, we have developed a 
consistent statistical description of seismicity using models of triggered
seismicity, which allows one to make quantitative
predictions of observables that can be compared with empirical data.
The simplest class of models of triggered seismicity combines the above
mentioned Gutenberg-Richter (i), Omori (ii), and productivity laws (iii)
which can be applied to a fractal spatial geometry of earthquake
epicenters (v) \cite{SaiSorpdf}. The fundamental physical ingredient is that
each earthquake can trigger other earthquakes
and an earthquake sequence results in
this model from the cascade of events triggered by
past earthquakes. The usual notions of foreshocks, mainshocks and aftershocks
lose their specificity as any earthquake can be triggered by previous
earthquakes and may trigger itself subsequent earthquakes. Here,
we continue our study of the general branching process, called the
Epidemic-Type Aftershock Sequence (ETAS) model of
triggered seismicity, introduced by Ogata in the present
form \cite{Ogata} and by Kagan and Knopoff in a slightly
different form \cite{KK81} and whose main average statistical
properties are reviewed in \cite{HS02}. This model has
been shown to constitute a powerful null hypothesis to
test against other models \cite{Ogata}. The ETAS model
belongs to a general class of branching processes
\cite{Athreya,Sankaranarayanan}. It can be viewed as the monofractal
approximation of the more general multifractal model of triggered
seismicity introduced recently in \cite{SO1_mf,OS1_mf}, which
derives from the physics of thermally activated rupture aided by stress.

The physical problem addressed here is the following. We start from the
empirical evidence \cite{H,HelmKaganJackson04} that small earthquakes
dominate or are at least equivalent collectively to 
large earthquakes in triggering other earthquakes. This can be seen
by combining the Gutenberg-Richter law (i) $\propto 10^{-bm}$ and the productivity law (iii)
$\propto 10^{\alpha m}$ to obtain the typical number $\propto 10^{-(b-\alpha)m}$ 
of events triggered by earthquakes of magnitude between $m$ and $m+1$.
With the empirical estimates of $b \approx 1$ and $0.8 \leq \alpha \leq 1$
together with the observation that triggered events seem to have magnitudes with only
weak or no relation with the magnitude of the triggering event 
(magnitude-independence law) \cite{HelmKaganJackson04}
(i.e., large earthquakes can be triggered by small events), 
this implies the perhaps surprising conclusion that large earthquakes
are triggered more by the swarm of small previous earthquakes than by preceding
large earthquakes. This stems from the observation that the number of small
earthquakes increases faster as their magnitude decrease than their productivity
decreases. The conclusion that small earthquakes dominate triggering is
thus intrinsically a collective effect. This picture, which emphasizes 
the collective organization of earthquakes or ``many-body'' view, can be contrasted with the
``one-body'' or few-body approach of R. Stein and co-workers
\cite{Stein1,Stein2} which focuses exclusively on how a few large earthquakes can promote
subsequent shocks at some sites and inhibit them in others. If indeed the small 
earthquakes dominate in the triggering of future events, this begs
to define how small ``small'' can be, since the smaller the earthquakes
the larger their triggering influence. 
The evidence that small earthquakes should dominate triggering is based on
the empirical statistics (i) and (iii) established for event 
magnitudes above magnitude 2 or 3 (depending on
the completeness of the studied catalogs). The question of how small ``small'' is
amounts to asking how far in the small magnitude range can the productivity law
and the magnitude-independence law be extrapolated. Because the
Gutenberg-Richter law (i) has been observed 
at such small scales as individual dislocation motions, we know for sure that 
there must be a lower ``ultra-violet'' cut-off magnitude $m_0$ at which 
the productivity of events of magnitude smaller than $m_0$ tapers off or
vanishes. Otherwise, the factor $\propto 10^{-(b-\alpha)m}$ would
diverges as $m \to -\infty$ (energy goes to zero). Is the ultra-violet
cut-off associated with an atomic scale for rupture? Or are other
relevant scales? This question has been addressed in two recent papers
by M. Werner and one of us \cite{SW1,SW2} within the framework of the
ETAS model. Consider a catalog complete for magnitudes above some
observational threshold $m_d$, i.e., all earthquakes with magnitudes $m
\geq m_d$ have been recorded but smaller earthquakes are not. Noting
that the magnitude $m_d$ of completeness of a seismic catalog is not in
general the same as the magnitude $m_0$ of the smallest triggering
earthquake, Ref.~\cite{SW1} showed that bounds for $m_0$ can be obtained
from quantitative fits to observed aftershock sequences. In addition, Ref.~\cite{SW2}
remarked that, in models of triggered seismicity and in their estimation
from empirical data, the detection threshold $m_d$ is commonly equated
to the magnitude $m_0$ of the smallest triggering earthquake. This
unjustified assumption neglects the possibility of shocks below the
detection threshold triggering observable events, a process which should
dominate according to our previous discussion. Ref.~\cite{SW2}
developed a mean field formalism within the ETAS model: by considering
the branching structure of one complete cascade of triggered events, the
catalog of observed events with magnitude above $m_d$ was shown to be
described by an effective ``renormalized'' ETAS model with its lower magnitude
cut-off equal to $m_d$ but with an apparent branching ratio $n_a$ (which is the
apparent fraction of aftershocks in a given catalog) and an apparent
background source $S_a$, due to the presence of smaller undetected
events capable of triggering larger events. This result is potentially
very important since it implies that previous estimates of the
clustering characteristics of seismicity may significantly underestimate
the true values: for instance, an observed fraction of $55\%$ of
aftershocks is renormalized into a true value of $75\%$ of triggered
events.

The object of the present paper is to extend the previous mean field
treatment to obtain the full earthquake statistics using the formalism
of generating probability function (GPF) already developed in
\cite{Saichevetal04,SaichSorl04,SSbath,SaiSorpdf}. In a sense, the
question addressed here is whether the ETAS model can be
renormalized onto itself by moving $m_0$ to $m_d>m_0$ (which can be seen
as a coarse-graining operation), that is, is there an effective ETAS
model with minimum magnitude $m_d$ and with renormalized parameters,
which describes the observed catalogs? Beyond its interest and 
application to earthquakes, this problem is relevant to a general
understanding of coarse-grained properties of marked branching processes, 
to which our formalism applies.

The organization of the paper is the following. Section \ref{2a} introduces
the general formulation of observable clusters of triggered events using the
generating probability functions (GPF). It also presents a simple
intuitive approximation which will be make rigorous in later sections.
Section \ref{2b} derives general relations for the effective branching rates
of observable events. Section \ref{2c} defines the ETAS model and recalls
its main useful properties. Section \ref{2d} introduces the GPF for
unobservable and observable aftershocks. Section \ref{mgmsls} gives the main
properties of the effective branching rates, which recover the previous
analysis of \cite{SW2} in a slightly different form. Section \ref{2f} explains
that the present approach and that of \cite{SW2} are equivalent physically
but with a different mathematical formulation. The
justification for introducing a physically equivalent but mathematically
different formulation here is that it
is more adapted to the calculations of the full statistics with the GPF formalism.
Section \ref{sec3} presents all our results on the statistics
of observable events in the ETAS branching model. Section \ref{3a} derives the
general equation governing the GPF of observable events. Section \ref{3b} use
the derivation of the previous section to give quantitative estimates for
the fraction of observable events. Section \ref{3c} discusses the approximation
of self-similarity, corresponding to a renormalization of the ETAS model onto
itself by the change from $m_0$ to $m_d$. This self-similarity amounts to say
that the statistics of observable events can be deduced entirely from the
statistics of all events under a simple renormalization of the average
branching ratio into an effective value. Section \ref{3d} derives the
implications of the self-similar approximation for the distribution of 
the numbers of observable events. Section \ref{3e} discusses the deviations
from self-similarity and identifies a correction in the form of a new
branching model, which gives rather small corrections to the previously
self-similar estimates. The last section concludes.

\section{Definition and properties of effective rate of observable aftershocks}

\subsection{General formulation of observable clusters \label{2a}}

In this section, we present the general formulation of generating
probability function (GPF) for marked branching processes with
an observational constraint. Recall that, for general branching processes
such as the ETAS model, the GPF formalism allows one to calculate the full
statistical properties. Here, the mark associated with an event
is its magnitude. The observation constraint is that only
events with magnitude $m \geq m_d$, where $m_d$ is the observation
threshold, are known, while the process produces events which can
have a lower magnitude, down to a lower triggering cut-off $m_0$.

To get a first feeling of how the observational constraint can be
taken into account in the GPF formalism, consider the case
of a large finite time window of size $\tau$ in which we
count the number of events and let us use the 
approach developed in \cite{SaichSorl04} for the statistics of the
number of such windowed events. The time windows are 
considered large if their size is significantly larger than the typical
life-time of the clusters, defined as the sequence of aftershocks,
triggered by single background event (see \cite{SaichSorl04} for
a discussion). In this limit, the statistics of the total
(observable and unobservable) number of events in a window of size $\tau$ 
is obtained from the generating
probability function (GPF) $\Theta_w(z;\tau)$, which obeys the 
following equation
\be
\Theta_w(z;\tau)= e^{\omega\tau [\Theta(z)-1]}~ . 
\label{1}
\ee
$\Theta(z)$ is the GPF of the number of all the aftershocks 
triggered by a given source, including the source event itself
and $\omega$ is the Poisson intensity of the background sources. 
We have shown \cite{Saichevetal04} that 
$\Theta(z)$ has the structure
\be
\Theta(z)= z G(z) ~,   \label{2}
\ee
where the factor $z$ to the left of $G(z)$ takes into account the contribution of
the background source, while the GPF $G(z)$ describes the statistics of the 
number of all aftershocks within a given cluster
In view of (\ref{2}), the GPF for finite time windows given by (\ref{1}) is the natural
generalization of the GPF obtained for the Poissonian
background events:
\be
\Theta_b(z;\tau)= e^{\omega \tau (z-1)}~ .
\ee

Now, the statistics of observable events requires to 
replace the GPF $\Theta(z)$ in (\ref{1}) by
the GPF $\Theta(z;m_d)$ of the number of aftershocks
(and their sources) whose magnitudes $m$ are larger than
the detection threshold $m_d$ to obtain 
\be
\Theta_w(z;m_d,\tau)= e^{\omega\tau [\Theta(z;m_d)-1]}~ . \label{3}
\ee
Note that some clusters might have no observable events
at all. This means that there is a non-zero probability
\be
p(m_d)= \Theta(z=0;m_d)\neq 0
\ee
that the cluster is completely unobservable. The 
complementary probability
\be
q(m_d)= 1- p(m_d)=1- \Theta(0;m_d)   \label{4}
\ee
is the probability that there is at least one observable
event (source or some aftershock) in the cluster under
inspection. In what follows, we refer to a cluster as
``observable,'' if it contains at least one observable event
(with magnitude $m \geq m_d$).
Accordingly, $q(m_d)$ defined in (\ref{4}) is the probability
that a cluster is observable; it is also the fraction of observable
clusters.

It is convenient to express the GPF $\Theta(z;m_d)$ in the form
\be
\Theta(z;m_d)= q(m_d) \tilde{\Theta}(z;m_d) + 1 - q(m_d)~ ,
\label{5}
\ee
where 
\be
\tilde{\Theta}(z;m_d)= {1 \over q(m_d)} \left[
\Theta(z;m_d)- \Theta(0;m_d)\right]   \label{6}
\ee
is nothing but the conditional GPF
of the number of observable events within observable clusters. 
This definition implies that it has the same structure
\be
\tilde{\Theta}(z:m_d)= z \tilde{G}(z;m_d)  ~, \label{7}
\ee
as that given by (\ref{2}) of the GPF $\Theta(z)$ of the total number of events
belonging to some cluster. Expression (\ref{7}) implies that one can treat
the observable event which comes first in time as the 
``observable source,'' and then interpret $\tilde{G}(z;m_d)$ as the GPF of its
observable aftershocks.

Substituting (\ref{5}) into (\ref{3}) obtains the following
representation for the GPF of the number of observable windowed events
\be
\Theta_w(z;m_d,\tau)= e^{\omega(m_d)\tau
[\tilde{\Theta}(z;m_d)-1]}~ , \label{8}
\ee
where
\be
\omega(m_d)= \omega q(m_d)    \label{9}
\ee
is the renormalized intensity of ``observable sources,''
which is the same as the intensity of observable clusters by definition.

There is a physically transparent way to estimate the
probability $q(m_d)$ that a cluster is observable. It is 
indeed always possible to represent $q(m_d)$ in the
form
\be
q(m_d)=q^+(m_d) Q(m_d)+ q^-(m_d) [1-Q(m_d)]~ .   \label{10}
\ee
Here, $q^+(m_d)$ (respectively $q^-(m_d)$)
is the probability that the cluster is observable
under the condition that its generating source is also
observable (respectively unobservable). In addition,
$Q(m_d)$ is the 
probability that the source is observable.
Obviously, we have
\be
q^+(m_d)\equiv 1~ , \qquad q^-(m_d)=1- p^-(m_d)~ ,   \label{11}
\ee
where $p^-(m_d)$ is the probability that all aftershocks
triggered by an unobservable event are unobservable. 
To estimate $p^-(m_d)$, we make the assumption
that each unobservable event 
either triggers
only one first-generation aftershock, with probability $\nu(m_d)$, or does not
trigger any aftershocks with the probability
$1-\nu(m_d)$. This approximation is quite reasonable, 
as can be seen from the application of the productivity law 
$\sim E^{a} \sim 10^{\alpha m}$ (with $a \approx 2/3$, $\alpha \approx 1$):
if an event of magnitude $m=7$ produces about $10^5$ observable events
on average, an event 
of magnitude $2$ triggers about $0.1$ events on average. In this example,
$\nu(m_d) \approx 0.1$ and $1-\nu(m_d) \approx 0.9$ and the error in 
neglecting the possibility for this event to trigger two aftershocks is
of order $0.01$. This error becomes even smaller for smaller unobservable sources.

Suppose additionally that the magnitudes of
the triggered aftershocks are statistically independent of each
other. Then, the probability that all
aftershocks, triggered by an unobservable event, are
unobservable, is given by
\be
p^-(m_d)\simeq \sum_{k=0}^\infty [1-\nu(m_d)] \nu^k(m_d)
[1-Q(m_d)]^k~ ,  
\ee
where $(1-\nu)\nu^k$ are the geometrical probability that an unobservable
event triggers $k$ aftershocks, while $(1-Q)^k$ is the
probability that they are all unobservable. After
summation, we obtain
\be
p^-(m_d)\simeq {1-\nu(m_d) \over 1- \nu(m_d) [1-Q(m_d)]}~.  \label{mgmlssswee}
\ee
Substituting this expression into (\ref{11}) and then (\ref{11}) into 
(\ref{10}), we obtain the probability that a cluster is observable
under the form
\be
q(m_d)\simeq {Q(m_d) \over 1- \nu(m_d) [1-Q(m_d)]}~ .
\label{12}
\ee

In the following, we obtain with the framework of the ETAS model, a
physically transparent relation for the probability
$\nu(m_d)$, which will allow us to obtain an accurate estimation of the
probability $q(m_d)$ given by (\ref{12}) and the corresponding
renormalized intensity $\omega(m_d)$ of ``observable
sources'' given by (\ref{9}). Specifically, the role of 
$\nu(m_d)$ is derived in expression (\ref{30}) below
for the GPF of first-generation aftershocks triggered by unobservable event.

\subsection{Effective observable aftershocks rate \label{2b}}

Before turning to the specifics of the ETAS model and its
statistics of observable events, it is useful
to discuss the properties of the average rate of general
branching processes. The results obtained in this section recover
those obtained in \cite{SW2} within a slighly different 
interpretation, that we present to be self-contained and to 
connect with the subsequent derivation of the full number statistics
in following sections.

It is well-known that the key parameter controlling the properties of
cascades of triggered events is the branching rate $n$,
which is nothing but the average number of first
generation aftershocks, where the average
is performed over all possible triggering event of arbitrary
magnitude. The cases $n<1$, $n=1$ and $n>1$ correspond
respectively to the sub-critical, critical and super-critical
regimes, with the first-two giving stationary time series in 
the presence of a stationary immigration of sources
and the later giving explosive time series with a positive probability
\cite{Athreya,Sankaranarayanan,HS02}.

In branching processes (of which the ETAS model is an example), 
we can use the representation that
each shock triggers independently its own aftershocks sequence (see
\cite{SW2} for a discussion on the two interpretations
in terms of decoupled branches used here or of collective triggering; the
two views are equivalent due to the linear sum over past events
and the conditional Poisson process formulation). The
independence between different branches allows us to obtain the 
average $\langle R\rangle$ of the total number of
events (mainshock itself and all its
offsprings over all geneations) triggered by an arbitrary mainshock as
\cite{HSn}
\be
\langle R\rangle= 1+n+ n^2+\dots ={1 \over 1-n} ~,
\label{13}
\ee
where $n^k$ is the contribution of the $k$-th generation of aftershocks. Thus, if 
the average number $\langle R\rangle$ of events per cluster is known, the
aftershocks rate can then be obtained as
\be
n={\langle R\rangle - 1 \over \langle R\rangle}~ .
\label{14}
\ee
This simple remark will be useful in the following to derive
an apparent or renormalized branching ratio $n(m_d)$ and test
its usefulness to describe the full number statistics.

The average number of
observable events, which are triggered by some arbitrary source, is
simply $n$ given by (\ref{14}) multiplied by the 
probability $Q(m_d)$ that an event is observable:
\be
Q(m_d)= \int_{m_d}^\infty P(m) dm~ , \label{15}
\ee
where $P(m)$ is the
probability density function (PDF) of their random
magnitudes (assumed to be the same for sources and all aftershocks). 
This gives
\be
\langle R\rangle(m_d)= \langle R\rangle Q(m_d)= {Q(m_d)
\over 1- n}~ .  \label{16}
\ee

Consider now the conditional average $\langle
\tilde{R}\rangle(m_d)$ of the number of observable events within some
observable cluster. It is simply given by
\be
\langle \tilde{R}\rangle(m_d)= {\langle R\rangle(m_d)
\over q(m_d)}= {Q(m_d) \over (1- n) q(m_d)}~ . \label{17}
\ee
This quantity $\langle
\tilde{R}\rangle(m_d)$ can not be derived as straightforwardly 
as the average number $\langle R\rangle$ over all events obtained with (\ref{13}).
Indeed, an observable cluster may result from an 
effective ``observable source'' which might
belong, for instance, to the $3$-th or even $7$-th generation
of the total aftershock sequence. Moreover, it seems
impossible to classify uniquely observable events of observable
clusters as belonging to observable aftershocks of first,
second or $k$-th generations. Therefore, it is not possible to use for
$\langle \tilde{R}\rangle(m_d)$ the
reasonings underlying relation (\ref{13}). Notwithstanding this
limitation, we can introduce an effective branching ratio
for observed clusters, based on a natural extension
of relation (\ref{14}). Let us thus define the effective
branching ratio of observable clusters as
\be
n(m_d)= {\langle \tilde{R}\rangle(m_d) - 1 \over \langle
\tilde{R}\rangle(m_d)}~ .
\ee
With (\ref{17}), this gives 
\be
n(m_d)= {Q(m_d)- (1-n) q(m_d) \over Q(m_d)}~ .  \label{mgmls}
\ee
Substituting in (\ref{mgmls}) the r.h.s. of equality (\ref{12}) yields
\be
n(m_d)\simeq {n -\nu(m_d) [1-Q(m_d)] \over 1-
\nu(m_d)[1-Q(m_d)]}~ ,   \label{18}
\ee
expressing the effective average aftershock rate via the probability
$Q(m_d)$ that a background event is observable and the
probability $\nu(m_d)$ that an unobservable background event
triggers some first-generation aftershock.

\subsection{Basic properties of the ETAS model \label{2c}}

To make further progress and in particular to calculate the 
probability $q(m_d)$ given by (\ref{12}) that a cluster is observable
and to obtain the effective branching rate $n(m_d)$ given
by (\ref{18}), we need to specify the properties of ETAS branching model.
The ETAS model is defined by the following rules. Each event of magnitude $m$
triggers a Poissonian sequence of aftershocks characterized by the
Poissonian GPF \cite{Saichevetal04}
\be
G_1(x;m|\kappa)= e^{\kappa \mu(m) (z-1)}~ ,  \label{19}
\ee
where $\kappa \mu(m)$ is the average number of first
generation aftershocks triggered by a mainshock of
magnitude $m$, $\kappa$ is a numerical constant and
$\mu(m)$ describes the so-called productivity law, i.e., the
dependence of the number of first generation aftershocks number on
the mainshock magnitude $m$. Previous empirical studies
have established that the productivity law is approximately exponential
\cite{H,HelmKaganJackson04}:
\be
\mu(m)= 10^{\alpha (m-m_0)}~ .   \label{promcmal}
\ee
with an exponent $\alpha$ in the range $0.8-1$. Here, $m_0$ is
the lower magnitude threshold below which events are supposed
not to be able to trigger any aftershock.
The ETAS model also uses the well-known Gutenberg-Richter (GR) law
for the PDF of earthquake magnitudes
\be
P(m)= b \ln(10) 10^{-b(m-m_0)}~ , ~~~{\rm with}~~ b \approx 1~, \label{20}
\ee
which are assumed to be independently drawn at each event occurrence.
Averaging the GPF defined by (\ref{19}) 
over all possible random source magnitudes $m$ weighted
by the GR distribution (\ref{20}), we obtain the GPF of first
generation aftershock numbers triggered by an arbitrary
source:
\be
G_1(z|\kappa)= F[\kappa (1-z)]~ , \label{21}
\ee
where
\be
F(x)= \gamma \int_1^\infty e^{-\kappa x y} {dy \over
y^{\gamma+1}}= \gamma y^\gamma \Gamma(-\gamma, y)~ ,
\qquad \gamma= {b \over \alpha}~ ,   \label{22}
\ee
and $\Gamma(-\gamma,y)$ is the incomplete Gamma function. In
the sequel, we shall use the following power law
expansion of the function $F(x)$ 
\be
F(x)\simeq 1- {\gamma \over \gamma -1}~ x+ \beta~
x^\gamma~ ,\quad \beta= \gamma \Gamma(-\gamma) = {\Gamma(2-\gamma) \over \gamma -1} .
\label{23}
\ee
Thus, for $\gamma \to 1+$, both coefficients of $x$ and $x^{\gamma}$
grow together. 

Our previous calculations have shown that this expansion is
very accurate for $\gamma \leq 1.25$, which is the relevant range
\cite{Saichevetal04,SaichSorl04,SSbath,SaiSorpdf}.

This expansion (\ref{23}) allows us to express
the main properties of the statistics of the number of aftershocks.
For this, let us substitute (\ref{23}) into (\ref{21}) to obtain
the corresponding approximate expression for the GPF of the number of first
generation aftershocks
\be
G_1(z|\kappa)\simeq 1-n(1-z)+ \beta \kappa^\gamma
(1-z)^\gamma~ ,  \label{24}
\ee
where
\be
n= \langle R_1\rangle ={\gamma \kappa \over \gamma- 1}
\label{25}
\ee
is the average aftershock branching ratio, i.e., the average
$\langle R_1\rangle$ of the total number of first generation
aftershocks triggered by an arbitrary source. Recall that
the last term $\beta \kappa^\gamma
(1-z)^\gamma$ in the r.h.s. of expression (\ref{24}) 
expresses the property that the distribution 
$\mathcal{P}_1(r|\kappa)$ of the total number of first
generation aftershocks triggered by an arbitrary source
has a power law tail
\be
\mathcal{P}_1(r|\kappa)\simeq {\gamma \kappa^\gamma \over
r^{1+\gamma}}~.   \label{26}
\ee
Expression (\ref{26}) is the leading asymptotic of the exact expression
\be
\mathcal{P}_1(r|\kappa)=  {1 \over r!} {d^r G_1(z|\kappa)
\over d z^r} \Big|_{z=0}= \gamma {\kappa^\gamma \over r!}
\Gamma(r-\gamma, \kappa)~ ,    \label{27}
\ee
corresponding to the exact GPF (\ref{21}) of the number of first generation
aftershocks. 
Ref.~\cite{Saichevetal04} has shown that 
the power law (\ref{26}) leads to a PDF of the total number of aftershocks
of all generations which are triggered by an arbitrary source,
which has a fatter tail $\sim 1/r^{1+(1/\gamma)}$, close to 
criticality $n \approx 1$ .

\subsection{Observable and unobservable aftershocks \label{2d}}

Let us now consider a different averaged GPF (\ref{19}) 
obtained by using a truncated GR law constrained
to unobservable earthquakes (with magnitudes $m$ between $m_0$ and $m_d$): 
\be
P^-(m|m_d)= {b \ln(10) 10^{-b(m-m_0)} \over 1- Q(m_d)}
H(m_d-m) H(m-m_0) ~,  \label{28}
\ee
where $H(x)$ is the unit step (Heaviside) function and $Q(m_d) = 10^{-b(m_d-m_0)}$
according to (\ref{15}) and (\ref{20}) is the probability for an earthquake
to be observable. Averaging expression
(\ref{19}) over all magnitudes weighted by $P^-(m|m_d)$ given by (\ref{28})
yields the GPF of the number of first-generation aftershocks triggered by an
unobservable event:
\be
G^-_1(z|\kappa, m_d)= {F[\kappa(1- z)]- Q(m_d) F[\kappa
\mu(m_d) (1-z)] \over 1-Q(m_d)}~ , \label{29}
\ee
Substituting the expansion (\ref{23}) in (\ref{29}) yields finally
\be
G^-_1(z|\kappa, m_d)\simeq 1- \nu(m_d)(1-z) + {\cal O}[(1-z)^2] ~,   \label{30}
\ee
where the coefficient $\nu(m_d)$ appears here from its definition 
as the probability that an unobservable background event
triggers some aftershock. The expansion (\ref{30}) at this linear order
for the GPF of first-generation aftershocks triggered by unobservable event
is equivalent to saying that an unobservable event can trigger at most
a single aftershock, in agreement with the approximation used to obtain
(\ref{mgmlssswee}) and (\ref{12}).

Expressions (\ref{23}) and (\ref{29}) thus yield
\be
\nu(m_d)= n {1- \rho(m_d) \over 1- Q(m_d)}~ ,   \label{31}
\ee
where
\be
\rho(m_d)= Q(m_d) \mu(m_d) = [\mu(m_d)]^{1-\gamma}   \label{rhoa,la}
\ee
describes the competition between the GR and productivity laws at the 
observational magnitude threshold $m_d$. Multiplying (\ref{31}) by 
the fraction $1-Q(m_d)$
of unobservable sources yields the average number $\langle
R^-_1\rangle$ of first generation aftershocks triggered
by an unobservable source. $\langle R^-_1\rangle$ can be interpreted as the 
branching rate $n^-(m_d)$ of first-generation aftershocks 
triggered by unobservable sources:
\be
\langle R^-_1\rangle(m_d)= n^-(m_d) = n [1- \rho(m_d)]~ .
\label{32}
\ee
Note that the GPF (\ref{30}) does not contain
a term of the form $\sim (1-z)^\gamma$ as in (\ref{24}), which was responsible for the
power law tail (\ref{26}) of the PDF of the total number of first generation
aftershocks. As a consequence, the tail of the PDF of first-generation aftershocks
triggered by unobservable sources is thinner than a power law. 
The power law tail (\ref{26}) is simply due to the
interplay between the productivity law (\ref{promcmal}) and the GR law 
(\ref{20}) for the sources. Now, constraining the source magnitudes to be 
smaller than $m_d$ truncates the GR law and thus the PDF of the number of
first-generation events.

Let us now turn to the statistics of first-generation
aftershocks triggered by observable sources. The
corresponding GPF is obtained by averaging (\ref{19}) over
all magnitudes weighted by the following modified GR law:
\be
P^+(m;m_d)= {P(m) \over Q(m_d)} H(m-m_d)= b \ln(10) 10^{-b
(m-m_d)} H(m-m_d)~ .
\ee
This leads to
\be
G^+_1(z|\kappa, m_d)= F[\kappa \mu(m_d)(1-z)]~ .  \label{33}
\ee
Note that expression (\ref{33}) differ from the GPF (\ref{21}) of 
the total number of first-generation events only through the
renormalization
\be
\kappa \qquad \rightarrow \qquad \kappa(m_d) = \kappa \mu(m_d) ~.    \label{34}
\ee
This allows us to interpret the 
average number of first-generation
aftershocks triggered by an arbitrary observable source
as an effective branching rate $n^+(m_d)$ equal to
\be
\langle R^+_1\rangle =n^+(m_d)= n \kappa(m_d) Q(m_d)= n
\rho(m_d)~ ,    \label{35}
\ee
where $\rho(m_d)$ is defined by (\ref{rhoa,la}). Not surprisingly,
the PDF of the number of first-generation aftershocks triggered by
observable background events has a power law tail, 
\be
\mathcal{P}^+_1(r|\kappa, m_d)=  {1 \over r!} {d^r
G^+_1(z|\kappa, m_d) \over d z^r} \Big|_{z=0}~,
\ee
analogous to (\ref{26}).

These different results are summarized in 
Figure 1 which shows the PDF's $\mathcal{P}^+_1(r|\kappa, m_d)$ and
$\mathcal{P}^-_1(r|\kappa, m_d)$ as a function of the number $r$
of events obtained from the exact relation (\ref{27}), 
for two different values of $\mu(m_d)$.

\subsection{Properties of effective aftershock rates \label{mgmsls}} 

We are now armed to discuss in the framework of the ETAS model the
properties of the probability $q(m_d)$ given by (\ref{12}) for a cluster
to be observable and the corresponding expression (\ref{18}) for the
effective branching rate $n(m_d)$ of observable clusters. 
Note again that the expansion (\ref{30}) at this linear order writes 
that an unobservable event can trigger at most
a single aftershock, in agreement with the approximation used to obtain
(\ref{mgmlssswee}) and (\ref{12}). This entitles us to 
substitute (\ref{31}) into (\ref{12}) and (\ref{18}) to obtain
\be
q(m_d)= {Q(m_d) \over 1- n[1- \rho(m_d)]}~ ,  \label{36}
\ee
and 
\be
n(m_d)= {n \rho(m_d) \over 1- n [1- \rho(m_d)]}~ ,
\label{37}
\ee
where $\rho(m_d)$ is defined by (\ref{rhoa,la}).
In the following sections, these two relations (\ref{36}) and (\ref{37})
will be derived from the exact equations obeyed by the GPF's
of the number of aftershocks over all generations.
In the mean time, let us discuss their properties and seismological implications.

Using the notations (\ref{32}) and (\ref{35}), we can rewrite the effective
rate (\ref{37}) of observable clusters in the form
\be
n(m_d)= {n^+(m_d) \over 1- n^-(m_d)}~,   \label{zxmmsl}
\ee
and interpret it as the rate $n^+(m_d)$ of aftershocks
triggered by observable sources, amplified by the impact of
aftershocks triggered by unobservable sources since the
denominator in (\ref{36}) and (\ref{37}) describes the influence of
aftershocks triggered by unobservable sources. 

First, notice that, in critical case $n=1$, we have
$n(m_d) \equiv 1$. Thus, the critical regime for all events is
also critical for observable events. In this case, the 
probability $q(m_d)$ that a cluster is observable is given by 
\be
q(m_d)= \mu^{-1}(m_d) \qquad (n=1)~ ,
\ee
and decreases as the observation threshold $m_d$ increases, which 
parallels the intensity of effective observable sources given by (\ref{9}).

Two cases are worth discussing. For
\be
n^-(m_d)= n [1- \rho(m_d)] \ll 1~,   \label{mgvkleas}
\ee
which corresponds to a negligible productivity of unobservable events, then
the impact of unobserved sources is small and
\be
q(m_d)\simeq Q(m_d)~ , \qquad n(m_d)\simeq n^+(m_d)= n
\rho(m_d)~ ,
\ee
as if all aftershocks, which are triggered by observable sources, were
observable.

In contrast, for
\be
\rho(m_d)\ll 1~,
\ee
we have
\be
q(m_d)\simeq {Q(m_d) \over 1- n}~ , \qquad n(m_d)\simeq {n
\rho(m_d) \over 1-n}~ ,
\ee
where the factor $1/(1-n)$, quantifying the impact of clusters
triggered by unobservable background events, becomes
predominant.

Expression (\ref{37}) can be rewritten as
\be
n(m_d) = {1 \over 1+  {1-n \over n}~[\mu(m_d)]^{\gamma -1}} ~.   \label{38}
\ee
Thus, for
\be
m_d-m_0 \ll \Delta^* \equiv {1 \over b-\alpha} \log_{10}\left({ n \over 1-n}\right)~ ,
\ee
the effective aftershocks rate is critical: $n(m_d)\simeq n$. For example, if
$n=0.9$, $b=1$ and $\alpha=0.8$ we have $\Delta^*\simeq
4.77$. Figure 2 (respectively 3) shows the dependence of 
the effective rate $n(m_d)$ as a function of 
$m_d-m_0$ (respectively $n$) for various $n$ (respectively $m_d-m_0$).

\subsection{Correspondence between the present formalism and Sornette-Werner
representation \cite{SW2} \label{2f}}

At this point, the astute reader will have noticed that the expression
(\ref{37}) for the effective rate of observable events of first-generation
is not the same as expression (10) of \cite{SW2}, which also gives
an apparent branching ratio denoted $n_a$ for observable aftershocks of the first
generation. Our present form (\ref{37}) for $n(m_d)$ departs from expression (10) of \cite{SW2}
for $n_a$ via the denominator, that is, by the fact that 
$n^-(m_d)$ defined in (\ref{mgvkleas}) is not zero. The two approaches
are actually equivalent as we now explain. Expression (10) of \cite{SW2}
defines an apparent branching rate $n_a$ as only due to observable sources while
$n(m_d)$ given by (\ref{37}) takes also into account 
the unobservable sources on observable aftershocks. In other words, 
$n(m_d)$ given by (\ref{37}) counts the effect of unobserved sources in the production
of first generation events and thus
describes the average number of first-generation daughters from
unobservable aftershocks which are themselves ``sources'' for the future
generations. In contrast, Sornette and Werner construct a representation 
in which the introduction of the observational cut-off
$m_d>m_0$ not only renormalizes $n$ into $n_a$ given by their equation (10)
but also introduces a renormalization of the spontaneous source rate \cite{SW2}: for each
real observable spontaneous sources, there are many apparent spontaneous
sources which result from the fact that an event triggered by an
unobservable previous aftershock is considered a spontaneous source
since one can not track its ancestor and can thus be counted as
spontaneous. The two approaches can thus be summarized as follows:
\ba
&& {\rm Sornette-Werner}:   \nonumber  \\
&& \{n , 1 ~{\rm spontaneous~source}\} \to  \{n_a = n \rho(m_d)~ ,~
S_a =N_{\rm obs} (n-n_a) ~{\rm spontaneous~ sources}\}~,
\ea
where $N_{\rm obs}$ is the total number of observed aftershocks;
\ba
&& {\rm present~work}:  \nonumber \\
&& \{n , 1 ~{\rm spontaneous~source}\} \to  \{n(m_d) ={n \rho(m_d) \over 1- n [1- \rho(m_d)]}~,~
 1 ~{\rm spontaneous~ source}\}~.
\ea
Note that the fraction $f_a = S_a/N_{\rm obs}$ of apparent sources among all
observed events given by expression (23) of \cite{SW2} can be written
\be
f_a = S_a/N_{\rm obs} = n - n_a = n - n \rho(m_d) = n^-(m_d)~,
\ee
where the last equality results from definition (\ref{mgvkleas}).
This provides a physically intuitive interpretation of $n^-(m_d)$.
Expression (\ref{37}) can thus be written
\be
n(m_d) = {n_a \over 1-f_a} =  n_a ( 1+ f_a + f_a^2 + ...)~.  \label{vmzmiq}
\ee
This formula clarifies completely the relationship between
Sornette-Werner's formation and the present one: 
the first term $n_a$ in the r.h.s. of (\ref{vmzmiq})
corresponds to the average number of daughters of first-generation due
to an observable initial source; The second term $n_a f_a$  corresponds to
the average number of daughters of first-generation which are
due to an apparent observable source which is triggered from a 
first-generation unobservable aftershock of the initial spontaneous source. The third
term $n_a f_a^2$  corresponds to the average number of daughters of
first-generation which are due to an apparent source 
which was itself triggered by an apparent source of a first-generation
unobserved aftershock of the initial spontaneous source; and so on...
This reasoning demonstrates that the two formulations
are physically equivalent, even though they have been obtained
by different physical arguments.

\section{Statistical description of observable events \label{sec3}}

Until now, we have explored some properties of the fraction $q(m_d)$ of
observable clusters and its corresponding effective
observable aftershock rate $n(m_d)$, using a physically
transparent but non-rigorous approach based on the
properties of first-generation aftershocks triggered by
observable and unobservable sources. In the following, we 
study the full statistical properties of 
observable events in large time window using the
GPF's of the number of events of all generations, belong to an
arbitrary cluster.

\subsection{Derivation of the GPF of observable events \label{3a}}

We start by the remark that the GPF of a single source of magnitude $m$, which 
takes into account the observability of the source, is equal to
\be
z^{H(m-m_d)}= 1+ (z-1) H(m-m_d)=
\begin{cases}
1~ , & m<m_d~ , \\
z~ , & m>m_d~ .
\end{cases}  \label{39}
\ee
Let us define $G(z;m,m_d)$ as the GPF of the number of
observable aftershocks of all generations which are triggered by 
a source (which can be observable or unobservable).
Multiplying $G(z;m,m_d)$ by (\ref{39}) yields the GPF $\Theta(z;m,m_d)$ of the number of
observable events triggered by a source of given magnitude $m$:
\be
\Theta(z;m,m_d)= z^{H(m-m_d)} G(z;m,m_d)~ .
\ee
Averaging this expression over all possible source magnitudes
weighted by the GR law (\ref{20}) gives the GPF
\be
\Theta(z;m_d)= \int_{m_0}^\infty \Theta(z;m,m_d) P(m) dm
\label{41}
\ee
of the total number of all observable events, which include
all observable sources and all their observable aftershocks of
all generations. $\Theta(z;m_d)$ can be expressed as
\be
\Theta(z;m_d)= G(z;m_d|m_0)+(z-1) G(z;m_d|m_d)   \label{40}
\ee
where
\be
G(z;m_d|x)= \int_x^\infty G(z;m,m_d) P(m) dm~ .   \label{42}
\ee
Thus, determining the GPF
$\Theta(z;m_d)$ requires to calculate the GPF $G(z;m,m_d)$ of the number of all
observable aftershocks of all generations belonging to the same cluster.
The later can be obtained by using the statistical independence of sources and
aftershocks magnitudes, which leads to replacing $z$ within the
exponential of the r.h.s. of (\ref{19}) by
$\Theta(z;m_d)$, which yields
\be
G(z;m,m_d)= e^{\kappa \mu(m) \left[ \Theta(z;m_d)-1
\right]}~ .   \label{43}
\ee
Substituting (\ref{43}) and (\ref{20}) into (\ref{42}) yields
\be
G(z;m_d|x)= Q(x) F(\kappa\mu(x)[1-\Theta(z;m_d)])~ . \label{mgjks}
\ee
Using this expression (\ref{mgjks}) to express the terms in the r.h.s.
of (\ref{40}) leads to the following equation determining the sought GPF
$\Theta(z;m_d)$:
\be
\Theta(z;m_d)= F(\kappa[1-\Theta(z;m_d)])+ (z-1) Q(m_d)
F(\kappa\mu(m_d)[1-\Theta(z;m_d)])~ .   \label{44}
\ee
For $m_d=m_0$ ($Q=\mu=1$) such that all events can be observed, this equation reduces to
the standard functional equation
\be
\Theta(z|\kappa)= z F(\kappa[1-\Theta(z|\kappa)])=z
G_1(\Theta(z|\kappa)|\kappa)~ ,    \label{45}
\ee
where $G_1(z|\kappa)$ is the GPF given by (\ref{21}) of the number
of first-generation aftershocks while
$\Theta(z|\kappa)=\Theta(z;m_0)$
is the GPF of the total number of event in a cluster. We make explicit
the dependence on the parameter $\kappa$ because it is going
to play a crucial role in the following discussion.

There is a physically natural
partition of the GPF $\Theta(z;m_d)$ given by (\ref{44}) according to
\be
\Theta(z;m_d)= G^-(z;m_d)+ z G^+(z;m_d)~ ,   \label{46}
\ee
where
\be
G^-(z;m_d)= F(\kappa[1-\Theta(z;m_d)])- Q(m_d)
F(\kappa\mu(m_d)[1-\Theta(z;m_d)])~ ,    \label{47}
\ee
describes the statistics of observable aftershocks triggered
by an unobservable event, while
\be
G^+(z;m_d)= Q(m_d) F(\kappa\mu(m_d)[1-\Theta(z;m_d)])~ .
\label{48}
\ee
describes the statistics of observable aftershocks
triggered by an observable event.

There are a few exact consequences of relations
(\ref{44})-(\ref{48}) which can now be obtained.
Consider the average number of events over all generations
of a given cluster, given by definition by
\be
\langle R\rangle(m_d)= {d\Theta(z;m_d) \over dz}\Big|_{z=1}~.
\ee
Using equation (\ref{44}), it is easy to show that it
satisfies the equation
\be
\langle R\rangle(m_d)= n \langle R\rangle(m_d)+ Q(m_d)~ ,
\ee
whose solution (\ref{16}) was already obtained from a direct
probabilistic argument. By construction, $\langle R\rangle(m_d)$ given
by (\ref{16}) is equal to the sum
\be
\langle R\rangle(m_d)= Q(m_d)+ \langle R\rangle^+(m_d)+
\langle R\rangle^-(m_d)   \label{49}
\ee
of the contributions of observable events and aftershocks, which are
triggered by both observable and unobservable events, with
\be
\langle R\rangle^\pm(m_d)= {d\Theta^\pm(z;m_d) \over
dz}\Big|_{z=1} ~.    \label{50}
\ee
It follows from (\ref{47}), (\ref{48}) and (\ref{50}) that
\be
\langle R\rangle^+(m_d)= n\rho(m_d) \langle R\rangle(m_d)=
n^+(m_d) \langle R\rangle(m_d) ~,  \label{51} 
\ee
and
\be
\langle R\rangle^-(m_d)= n[1-\rho(m_d)] \langle
R\rangle(m_d)= n^-(m_d) \langle R\rangle(m_d)~.
\label{52}
\ee
These two relations confirm the physical meaning of 
the rates $n^\pm(m_d)$ defined in (\ref{32}) and (\ref{35}),
which quantify the relative impact of aftershocks
triggered by observable versus unobservable events. 
Expressions (\ref{49}), (\ref{51}) and (\ref{52})
show that the rates $n^\pm(m_d)$ are the fractions of
aftershocks of all generations which are triggered by 
observable ($^+$) versus unobservable ($^-$) events.

Figure 4 plots these two rates $n^\pm(m_d)$ as a function of 
$m_d-m_0$ for $\alpha=0.8$, $b=1$, $n=0.9$, showing that
the impact of unobserved events may easily dominate
for quite reasonable values of the model parameters.

\subsection{Fraction of observable clusters \label{3b}}

One of the key parameters governing the statistics of
windowed observable events is the fraction $q(m_d)$ defined by 
(\ref{4}) of observable
clusters. Equation (\ref{44}) allows us to calculate it exactly.
Indeed, it is easy to show that expression (\ref{44}) implies that 
$q(m_d)$ is solution of the equation
\be
q(m_d)=1- F[\kappa q(m_d)]+ Q(m_d)F[\kappa\mu(m_d)q(m_d)]~. 
\label{53}
\ee
Noticing that $Q(m_d)=[\mu(m_d)]^{-\gamma} \equiv \mu^{-\gamma}$, 
we can rewrite (\ref{53}) in the form
\be
\Psi[q(m_d)]=0~,    \label{54}
\ee
where
\be
\Psi(x)= 1-x- F(\kappa x)+ \mu^{-\gamma} F(\kappa \mu x)~. 
\label{55}
\ee
A good approximate solution of (\ref{54}) can be obtained
by substituting the polynomial approximation (\ref{23}) for $F$ to obtain
\be
\Psi(x)\simeq \mu^{-\gamma}- x[1-n (1-\mu^{1-\gamma})]~ .
\ee
The corresponding solution of {\ref{54}) then reads
\be
q(m_d)= {1 \over n \mu+ (1-n) \mu^\gamma}~ ,   \label{56}
\ee
which is equivalent to expression (\ref{36}) derived above using
an intuitive nonrigorous
reasoning. In contrast, expression (\ref{56}) and thus (\ref{36})
is now obtained as an approximate
solution of the exact equation (\ref{53}). The accuracy of this 
approximation (\ref{56}) (or (\ref{36})) can thus be checked by comparing 
it with the numerical solution of the exact equation (\ref{53}).
Correlatively, this also directly check the quality of expression
(\ref{37}). Figure 5 shows the ratio of the approximation (\ref{56})
divided by the numerical solution of the exact equation (\ref{53}),
as a function of $m_d-m_0$ for $n=0.9, b=1$ and three values of $\alpha=0.7, 0.8, 0.9$.
One can observe that the quality of the approximation (\ref{56}) improves
as $\alpha$ gets closer to $1$.

\subsection{Self-similarity of the statistics of observable events \label{3c}}

We now have the tools to calculate the conditional
GPF $\tilde{\Theta}(z;m_d)$ defined by (\ref{6})
of the total number of observable events within an
observable cluster. Substituting (\ref{5}) into (\ref{44}) yields the
equation for $\tilde{\Theta}(z;m_d)$:
\be
\varphi(\tilde{\Theta};m_d)= z
F(\kappa(m_d)(1-\tilde{\Theta}))~ ,   \label{57}
\ee
where
\be
\kappa(m_d)= \kappa \mu(m_d) q(m_d) ~,  \label{58}
\ee
and
\be
\varphi(x;m_d)= {\Psi(q(m_d)(1-x)) \over Q(m_d)}~ .
\label{59}
\ee
Definition (\ref{55}) and equation (\ref{54}) imply that the
following identities are true
\be
\varphi(0;m_d)\equiv 0~ , \qquad \varphi(1;m_d)\equiv 1~ .
\label{60}
\ee
Using (\ref{55}) and the approximate expression (\ref{23}),
we obtain the linear approximation
\be
\varphi(x;m_d)\simeq \varphi_1(x)~ , \qquad \varphi_1(x)= x~ ,   \label{61}
\ee
which is consistent with (\ref{60}).
Then, substituting (\ref{61}) into (\ref{57}) yields an approximate
equation for the GPF $\tilde{\Theta}(z;m_d)$:
\be
\tilde{\Theta}(z;m_d) \simeq z
F(\kappa(m_d)[1-\tilde{\Theta}(z;m_d)])~ .   \label{62}
\ee

We can check the accuracy of the linear approximation (\ref{61}),
and thus its consequence for $\tilde{\Theta}(z;m_d)$ given by 
(\ref{62}) by comparing the linear function $\varphi_1(x)= x$
of (\ref{61}) with the exact one given by (\ref{57}).
Figure 6 shows the difference
\be
\Delta_1(x;m_d)=\varphi(x;m_d)- x~ ,   \label{63}
\ee
where $\varphi(x;m_d)-$ is given by (\ref{57}),
as a function of the variable $x$, for $n=0.9, \gamma=1.25$ and $\gamma=1.1$
and several values of $m_d-m_0$. This figure confirms
the good accuracy of the linear approximation. The
two following subsections will extract the consequence of
this formulation for the distribution of aftershock numbers
and will quantify the impact of the next order correction to the
linear approximation (\ref{61}).

Note that equation (\ref{62}) coincides, after the 
application of the renormalization
(\ref{34}) where $\kappa(m_d)$ is now given by expression (\ref{58}), with
the equation (\ref{45}) for the GPF $\Theta(z|\kappa)$ of the total number
of events within an arbitrary cluster. This has an important
consequence for the physical understanding of seismicity according to
the ETAS model: as long as the linear approximation (\ref{61}) is
applicable, the statistics of the number of observable events within
observable clusters is identical, up to the renormalization
(\ref{34}), to the statistics of the total number
of events within an arbitrary cluster in which all events can be observed.  

This can be restated as the following
self-similar property for the statistical properties of observable clusters:
\be
\tilde{\Theta}(z;m_d)\simeq \Theta(z|\kappa(m_d))~ .
\label{64}
\ee
This self-similarity property means that the statistics of observable
windowed events within large time windows is identical after the
correspondence
\be
\omega\quad \rightarrow\quad \omega(m_d)=\omega q(m_d)~ , \qquad n \quad
\rightarrow \quad n(m_d)={\gamma
\kappa(m_d) \over \gamma-1}~ ,   \label{65}
\ee
to the statistics of the total number of windowed events. 
The effective branching rate $n(m_d)$ defined in (\ref{65})
coincides, using the expression (\ref{58}) for $\kappa(m_d)$ 
and (\ref{56}) for $q(m_d)$, with expression (\ref{37}) 
that we have previously obtained for the effective rate of
observable aftershocks. This shows again that the intuitive
probabilistic reasoning of section \ref{mgmsls} on 
effective aftershock rates is equivalent to the linear approximation
(\ref{61}) for the GPF. As we are going to probe in greater depth,
this suggests that the self-similar property (\ref{64}) is a
resilient and general feature of the ETAS model.

\subsection{Distribution of the number of observable events \label{3d}}

We now derive the consequences of the above results for the
distribution of the numbers of observable events.

Let us denote by $\tilde{\mathcal{P}}(r;m_d)$ the probability
corresponding to the GPF $\tilde{\Theta}(z;m_d)$ defined by (\ref{6}),
that there are $r$ observable events in a given observable
cluster. Similarly, we denote $\mathcal{P}(r;m_d)$ the distribution of
the numbers of observable events within an arbitrary cluster
corresponding to the GPF $\Theta(z;m_d)$. The two
GPF $\tilde{\Theta}(z;m_d)$ and $\Theta(z;m_d)$ are linked through
equation (\ref{5}). It follows from (\ref{5}) and (\ref{62}) that, 
within the domain of application of the linear approximation (\ref{61}),
these two probabilities can be expressed in terms of the probability
$\mathcal{P}(r|\kappa)$ of the total number of events of an arbitrary cluster
via the following self-similar relations
\be
\tilde{\mathcal{P}}(r;m_d)\simeq
\mathcal{P}(r|\kappa(m_d))~ , \quad
\mathcal{P}(r;m_d)\simeq q(m_d)
\mathcal{P}(r|\kappa(m_d))~ , \quad r\geqslant 1~ .
\label{66}
\ee
Thus, the statistical properties of observable
events are known from those of the all events via the scaling
relations (\ref{66}) (within the linear approximation (\ref{61}) of the GPF).
The self-similar properties (\ref{64}) and (\ref{66})
mean that the ETAS model is renormalized onto itself under 
the transformation $m_0 \to m_d$, with just a renormalization 
from $\kappa$ to $\kappa(m_d)$ and, as a consequence, a renormalization of the 
branching ratio from $n$ to $n(m_d)$. Our present analysis thus confirms
for the full statistical properties the results obtained previously, based
solely on the average seismic rates \cite{SW2}.

The statistics properties of the total number of events in 
individual aftershock clusters (without the constraint of observability) 
has been derived in our previous paper \cite{Saichevetal04,Saichevetal04}.
Therefore, we just need to recall briefly some of its key properties
which are useful for understanding the statistics of
observable events.

Recall that the probability density $\mathcal{P}(r|\kappa)$ 
is given by the Cauchy integral \cite{SaiSorpdf} 
\be
\mathcal{P}(r|\kappa)= {1 \over 2\pi i r}
\oint\limits_{\mathcal{C}} {d \Theta(z|\kappa) \over z^r}~ , 
\label{67}
\ee
where $\mathcal{C}$ is sufficiently small contour
enveloping the origin $z=0$, and $\Theta(z|\kappa)$ is
solution of the functional equation (\ref{45}). The main difficulty in the
calculation of the integral (\ref{67}) is that the GPF
$\Theta(z|\kappa)$ is defined only implicitly, via the
solution of equation (\ref{45}). To overcome this obstacle, we perform
a change of variable and use the new integration variable $y=
\Theta(z|\kappa)$. It follows from (\ref{45}) that
\be
z= {y\over G_1(y|\kappa)}~ ,
\ee
which yields the explicit integral for
$\mathcal{P}(r|\kappa)$:
\be
\mathcal{P}(r|\kappa)= {1 \over 2\pi i r}
\oint\limits_{\mathcal{C}'} G_1^r(y|\kappa){d y \over y^r}~ , 
\label{68}
\ee
where $\mathcal{C}'$ is some small contour in the
complex plane $y$ enveloping the origin $y=0$.

It is interesting to point out that relation (\ref{68}) 
has an intuitive probabilistic interpretation, as it can be 
expressed as 
\be
\mathcal{P}(r|\kappa)= {1 \over r}\, \text{Pr\,}(Y_r= r-1)~ ,   
\label{69}
\ee
where
\be
Y_s= \sum_{k=1}^s U_k ~,  \label{70}
\ee
and $\{U_1,U_2,\dots\}$ are mutually 
independent random integers with GPF $G_1(z|\kappa)$ given by
(\ref{21}) with (\ref{24}). In the relevant regime for 
earthquakes for which, probably, $1<\gamma<2$, and for $r\gg 1$, the PDF of 
the sum (\ref{70}) tends asymptotically to
\be
\text{Pr\,}(Y_r= s)= {1 \over (\nu r)^{1/\gamma}}
\ell_\gamma \left( {s- n r \over (\nu
r)^{1/\gamma}}\right)~, \qquad \nu= -\kappa^\gamma
\Gamma(1-\gamma)~ ,   \label{71}
\ee
where $\ell_\gamma(x)$ is the stable L\'evy distribution such
that its two-sided Laplace transform is equal to
\be
\int_{-\infty}^\infty \ell_\gamma(x) e^{-u x} dx=
e^{u^\gamma}~ .
\ee
This L\'evy distribution has the following properties 
\be
\ell_\gamma(x)\sim {x^{-\gamma-1} \over \Gamma(-\gamma)}
\quad (x\to \infty)~ , \qquad \ell_\gamma(0)= {1 \over
\gamma \Gamma(1-1/\gamma)}~ .   \label{72}
\ee
One can calculate $\ell_\gamma(x)$ numerically for any
value $1<\gamma<2$ using, for instance, the following integral
representation
\be
\ell_\gamma(x)= {1 \over \pi} \int_0^\infty \exp\left[
-u^\gamma+ u x \cos \left({\pi \over \gamma}\right)
\right] \sin\left[u x \sin\left({\pi \over \gamma}\right)+
{\pi \over \gamma}\right] du ~ .
\ee

The asymptotic expression for the probability (\ref{69}) 
corresponding to (\ref{71}) is
\be
\mathcal{P}(r|\kappa)\simeq {1 \over r (\nu
r)^{1/\gamma}}\, \ell_\gamma \left( {r(1-n)-1 \over (\nu
r)^{1/\gamma}}\right)\qquad (r\gg 1)~ .  \label{73}
\ee
When the average branching ratio $n$ defined by (\ref{14})
is close to $1$,Eq.~(\ref{73}) with (\ref{72})
predicts the existence of two characteristic
power laws  for the probability $\mathcal{P}(r|\kappa)$:
\be
\mathcal{P}(r|\kappa)\sim r^{-1- 1/\gamma} \qquad (r\ll
r^*) ~,   \label{74}
\ee
and
\be
\mathcal{P}(r|\kappa)\sim r^{-1-\gamma} \qquad (r\gg r^*)~,
\label{75}
\ee
where
\be
r^*= \nu^{1/(\gamma-1)} \left( {1 \over
1-n}\right)^{\gamma/(\gamma-1)}~ .   \label{76}
\ee
The power law (\ref{75}) reflects the intrinsic distribution
of the number of first-generation aftershocks given by
relation (\ref{26}), while the heavier power law tail (\ref{74}) reflects
the effects of cascades over many generations in the branching 
aftershocks triggering process \cite{Saichevetal04}.
See figure 2 in Ref.~\cite{Saichevetal04} for a visualization
of the two power laws (\ref{74}) and (\ref{75}) and their cross-over.

Then, substituting in (\ref{76}) the effective branching rate $n(m_d)$ for observable
clusters, we obtain the dependence of the cross-over value $r^*(m_d)$ 
separating the two power laws (\ref{74}) and (\ref{75}) for the statistics
of the number of observable events, as a function of 
the threshold magnitude $m_d-m_0$. Figure 7 shows $r^*(m_d)$ as a function of 
$m_d-m_0$ for $\gamma=1.25$ and several values of $n$.
One can observe a fast decrease of $r^*(m_d)$ with $m_d-m_0$, which
implies that increasing the observation magnitude threshold $m_d$ amounts
to deviate more and more from criticality, as shown also directly
in Figure 2.

\subsection{Deviations from self-similarity \label{3e}}

All above results on the self-similarity of the
statistics of observable events expressed by relations
(\ref{64}) and (\ref{66}) can be viewed as the consequence of
the linear approximation (\ref{61}). It is thus important
to explore how strong can be the deviations from 
self-similarity resulting from the properties of the 
exact equation (\ref{57}) for the GPF
$\tilde{\Theta}$ of the number of observable events within an
observable cluster. In this goal, we rewrite equation (\ref{57}) in
the form
\be
\tilde{\Theta}= z \tilde{G}_1(\tilde{\Theta};m_d)~,
\label{77}
\ee
where
\be
\tilde{G}_1(z;m_d)= G_1(z|\kappa(m_d) g(z;m_d)~,  \label{78}
\ee
and
\be
g(z;m_d)= {z \over \varphi(z;m_d)}~ .   \label{79}
\ee
One can interpret (\ref{77}) and (\ref{78}) as describing some
branching process such that the GPF of the number of first-generation
aftershocks is equal to $\tilde{G}_1(z;m_d)$. In other
words, the random number $R_1(m_d)$ of first-generation
aftershocks in this new branching model is equal to the sum of 
the two statistically independent random integers
\be
R_1(m_d)= U(m_d)+ V(m_d)~ ,
\ee
where $U(m_d)$ has the self-similar GPF $G_1(z|\kappa(m_d)$,
while $V(m_d)$ has the GPF $g(z;m_d)$.

Using (\ref{68}), (\ref{69}) and (\ref{78}), we immediately obtain
the exact integral representation of the PDF of the number
of observable events in an observable cluster:
\be
\mathcal{P}(r;m_d)= {1 \over 2\pi i r}
\oint\limits_{\mathcal{C}'} G_1^r[z|\kappa(m_d)]
g^r(y;m_d){d y \over y^r} ~ . \label{80}
\ee
Its corresponding probabilistic representation reads
\be
\tilde{\mathcal{P}}(r;m_d)= {1 \over r}\, \text{Pr\,}(Y_r=
r-1)~ ,    \label{81}
\ee
where
\be
Y_s= \sum_{k=1}^s (U_k+V_k)~ .   \label{82}
\ee
The random integers $\{U_1,U_2,\dots\}$ have the GPF
$G_1[z|\kappa(m_d)]$, while the random numbers $\{V_1, V_2,\dots\}$ are
random integers which are 
mutually statistically independent from each other (and from
$U$'s) with the GPF $g(y;m_d)$ given by (\ref{79}).
As the transformation $m_0 \to m_d$ is equivalent to a decimation
step in the language of the renormalization group (see
\cite{wilsonkogut} and \cite{Wilson} as well as Chapter 11 of \cite{sorbook}
for pedagogical introductions), the random variables $\{V_1, V_2,\dots\}$
with their GPF $g(y;m_d)$ correspond to a 
new relevant direction (branching process different from ETAS) 
in the space of branching processes.

To obtain the statistical properties of the random integers $V$,
consider the quadratic approximation of the function
$\varphi(x;m_d)$ defined by (\ref{59}):
\be
\varphi(x;m_d)\simeq \varphi_2(x;m_d)~ , \qquad
\varphi_2(x;m_d)=x +\eta(m_d) x(1-x)~ .   \label{83}
\ee
This second-order approximation is again
consistent with the identities (\ref{60}). Here,
\be
\eta(m_d)= 4 \Delta_1(1/2;m_d)~ ,  \label{84}
\ee
where $\Delta_1$ is defined in (\ref{63}).
Figure 8 shows the difference
\be
\Delta_2(x;m_d)= \varphi(x;m_d)- \varphi_2(x;m_d)~ ,
\label{85}
\ee
as a function of $x$ for different values of $m_d-m_0$ 
for the same parameters as for Figure 6.
$\gamma=1.1$ and $1.25$. Comparison between Figures 6 and 8 
demonstrate the large improvement from the linear to the 
quadratic approximation (\ref{83}).

Substituting (\ref{83}) into (\ref{79}) yields the approximate GPF of
the auxiliary random integers $V$ as
\be
g(z;m_d)= {1 \over 1+\eta(m_d) (1-z)}~ .  \label{86}
\ee
This GPF means that $V$ has a geometric distribution with the
following average and variance
\be
\langle V\rangle= \eta~ , \qquad \sigma^2_V= \eta (\eta+1)~ .
\ee
Thus, if $\eta\ll 1$, the random variable $V$ has a
small impact on the statistics of the number of observable events. In
this case, we obtain the leading asymptotical contribution of
the variable $V$ to the statistics of observable events by using
a power law expansion for the GPF
$\tilde{G}_1(z;m_d)$, similar to (\ref{24}):
\be
\tilde{G}_1(z;m_d)\simeq 1+ \tilde{n}(m_d) (1-z) + \beta
\kappa^\gamma(m_d) (1-z)^\gamma~ ,   \label{87}
\ee
where
\be
\tilde{n}(m_d)= n(m_d)+ \eta(m_d)~ .    \label{88}
\ee
Expression (\ref{87}) shows that the main contribution of the random
integer $V$ resulting from the first-order correction to the 
linear approximation (\ref{61}) is to introduce a small shift
(for $\eta\ll 1$) equal to
$\eta(m_d)$ to the effective branching rate $n(m_d)$ obtained within
the linear approximation (\ref{61}). 
Figure 9 shows the dependence of this shift
$\eta(m_d)$ as a function of $m_d-m_0$ for different values of $\gamma$.
Since $n(m_d)$ is typically in the range $0.5-1$, this shows that
the corrections are no more than about $10\%$ in the value of the
effective branching rate for observable events.

\section{Conclusion}

We have shown that, to a good approximation, the ETAS model
is renormalized onto itself under what amounts to 
a decimation procedure $m_0 \to m_d$, with just a renormalization of the 
branching ratio from $n$ to an effective value $n(m_d)$. 
Our present analysis thus confirms,
for the full statistical properties, the results obtained previously
in Ref.~\cite{SW2}, based
solely on the average seismic rates (the first-order moment of the
statistics). However, our analysis also demonstrates that this
renormalization is not exact, as there are small corrections
which can be systematically calculated, in terms of additional
contributions that can be mapped onto a different branching model
(a new relevant direction in the language of the renormalization group).
However, for practical applications, due to the strong stochasticity
of the ETAS branching model, these deviations from exact self-similarity
will be difficult to observe. This justifies the standard
procedure in statistical parameter 
estimations of using the ETAS model with magnitude cut-off $m_d$
even if $m_d$ is an artificial detection threshold with no
physical meaning for the triggering process. However, 
our results, which confirm by and large the conclusions
of Ref.~\cite{SW2}, show that the values of the 
branching ratio (or average rate of generation of first-generation
aftershocks) recovered by such statistical estimations is not
the ``true'' one, but a effective or renormalized value. 
Thus, conclusions of the properties of aftershock clusters
has to be re-examined in this light: echoing the
main conclusion of Ref.~\cite{SW2}, `` previous estimates of the
clustering characteristics of seismicity may significantly underestimate
the true values.''

{\bf Acknowledgments:} We thank M. Werner for stimulation discussions.
This work is partially supported
by NSF-EAR02-30429, and by the Southern California
Earthquake Center (SCEC) SCEC is funded by NSF
Cooperative Agreement EAR-0106924 and USGS Cooperative
Agreement 02HQAG0008. The SCEC contribution number for
this paper is X.

{}

\clearpage

\begin{quote}
\centerline{
\includegraphics[width=14cm]{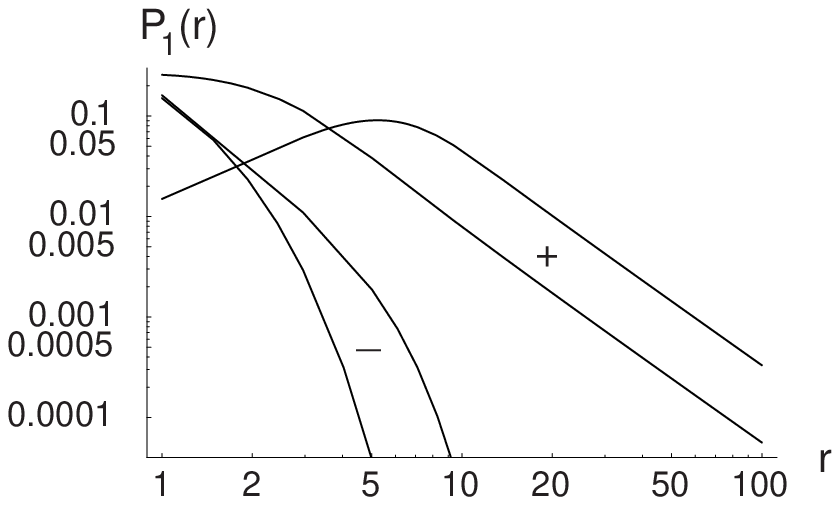}}
{\bf Fig.~1:} \small{Dependence of the PDF's 
$\mathcal{P}^+_1(r|\kappa, m_d)$ and
$\mathcal{P}^-_1(r|\kappa, m_d)$ as a function of number $r$ for $\gamma=1.1$, $n=0.9$
and for $\mu(m_d)=10$ and $50$, illustrating the presence of a power law
tail $\sim r^{-\gamma-1}$ for first-generation aftershocks triggered by
observable sources and of fast decaying tails for first-generation 
aftershocks triggered by unobservable sources (of magnitude less than $m_d$).
 }
\end{quote}

\clearpage

\begin{quote}
\centerline{
\includegraphics[width=14cm]{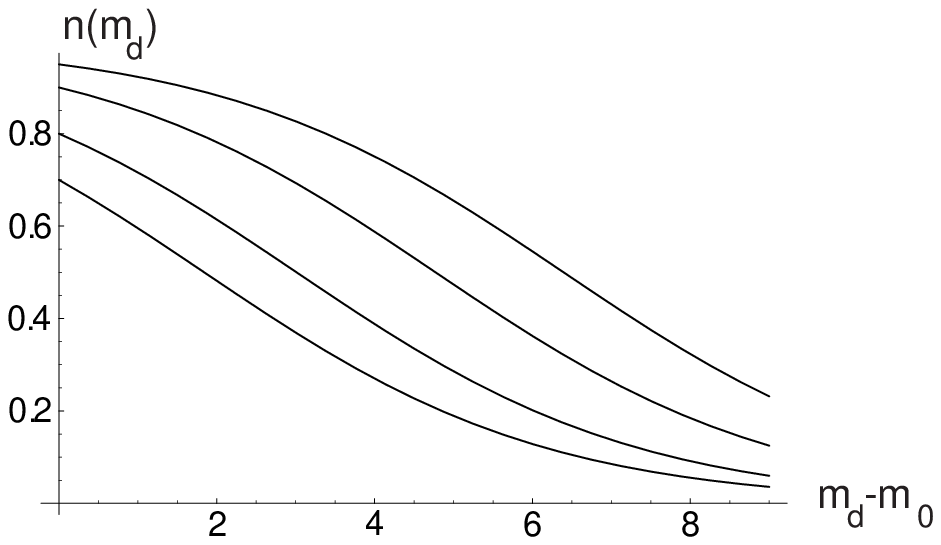}}
{\bf Fig.~2:} \small{Dependence of the effective rate $n(m_d)$
given by (\ref{37}) and (\ref{38}) for $\alpha=0.8$ and $b=1$ ($\gamma=1.25$)
for different value of $n$: $n=0.7;0.8;0.9;0.95$ from bottom to top.}
\end{quote}

\clearpage

\begin{quote}
\centerline{
\includegraphics[width=14cm]{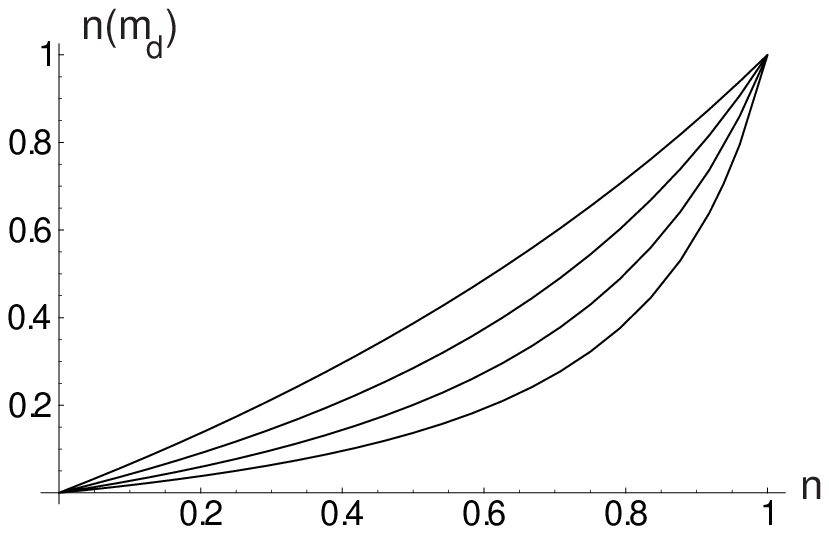}}
{\bf Fig.~3:} \small{Dependence of effective rate
$n(m_d)$ given by (\ref{37}) and (\ref{38}) as a function of the 
branching ratio $n$ of all first-generation events, for $\alpha=0.8$
and $b=1$ and several values of $m_d-m_0$: $m_d-m_0=1;2;3;4$ from bottom to top. }
\end{quote}

\clearpage

\begin{quote}
\centerline{
\includegraphics[width=14cm]{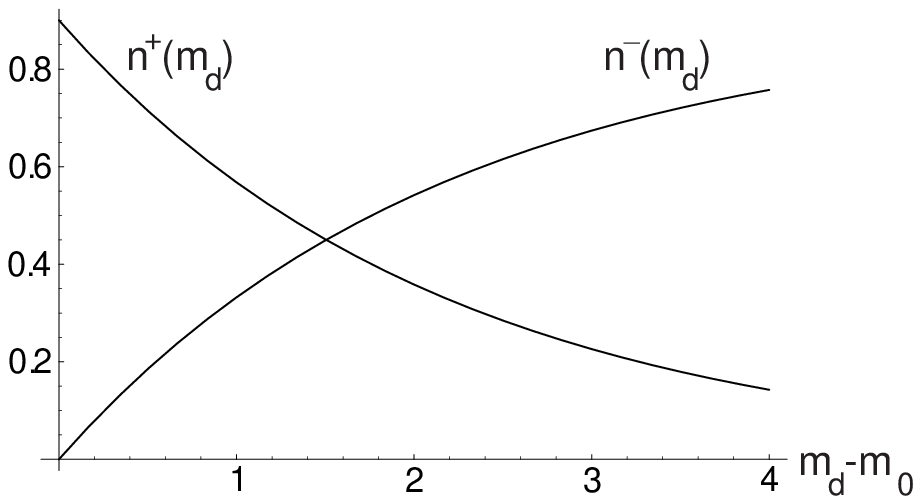}}
{\bf Fig.~4:} \small{Dependence of the rates $n^\pm(m_d)$ quantifying
the relative impact of aftershocks triggered by observable ($^+$)
versus unobservable ($^-$) events,
as a function of $m_d-m_0$, for $\alpha=0.8$, $b=1$, and $n=0.9$. }
\end{quote}

\clearpage

\begin{quote}
\centerline{
\includegraphics[width=14cm]{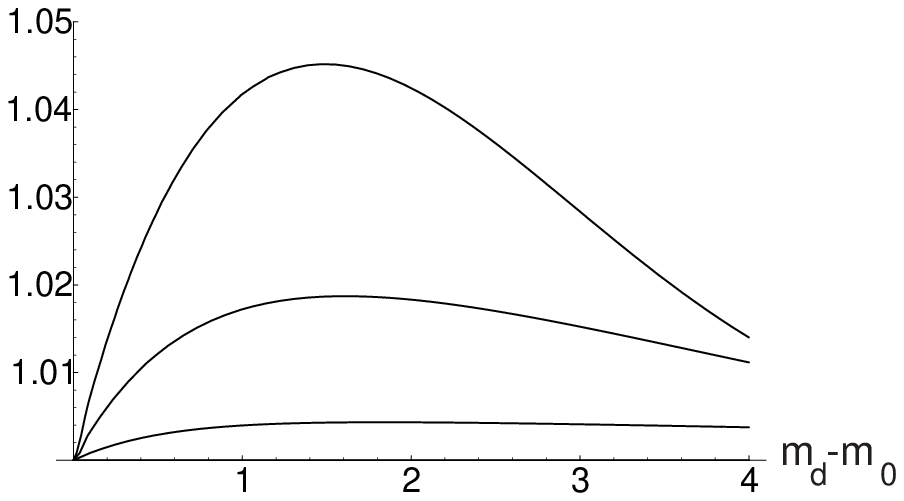}}
{\bf Fig.~5:} \small{Dependence of the ratio of the approximation (\ref{56})
divided by the numerical solution of the exact equation (\ref{53}),
as a function of $m_d-m_0$, demonstrating the good 
accuracy of the approximate expression (\ref{56}), for $n=0.9$,
$b=1$: $\alpha=0.7;0.8;0.9$ from top to bottom. }
\end{quote}

\clearpage

\begin{quote}
\centerline{
\includegraphics[width=14cm]{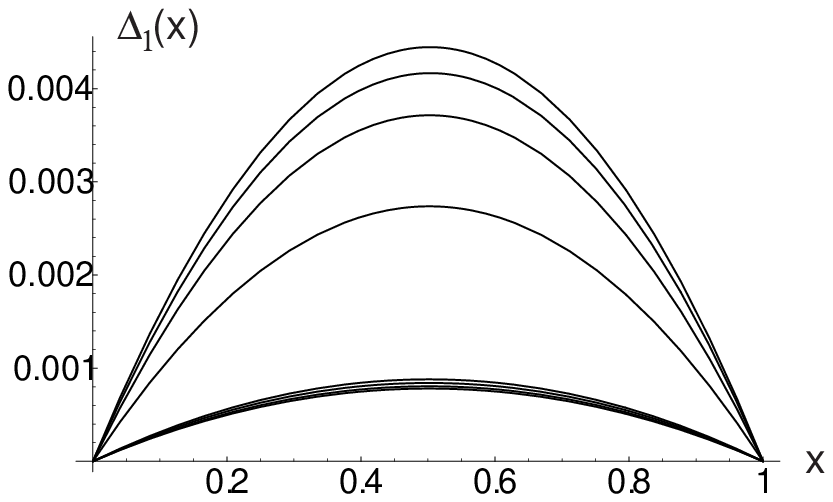}}
{\bf Fig.~6:} \small{Dependence of the difference $\Delta_1(x;m_d)$
given by (\ref{63}) as a function of the variable $x$, for $n=0.9, \gamma=1.25$
and several values of $m_d-m_0 =1;2;3;4$ (four upper curves from top to bottom).
The group of almost undistinguishable curves at the bottom of the graph
corresponds to $n=0.9, m_d-m_0 =1;2;3;4$ and $\gamma=1.1$.
 }
\end{quote}

\clearpage

\begin{quote}
\centerline{
\includegraphics[width=14cm]{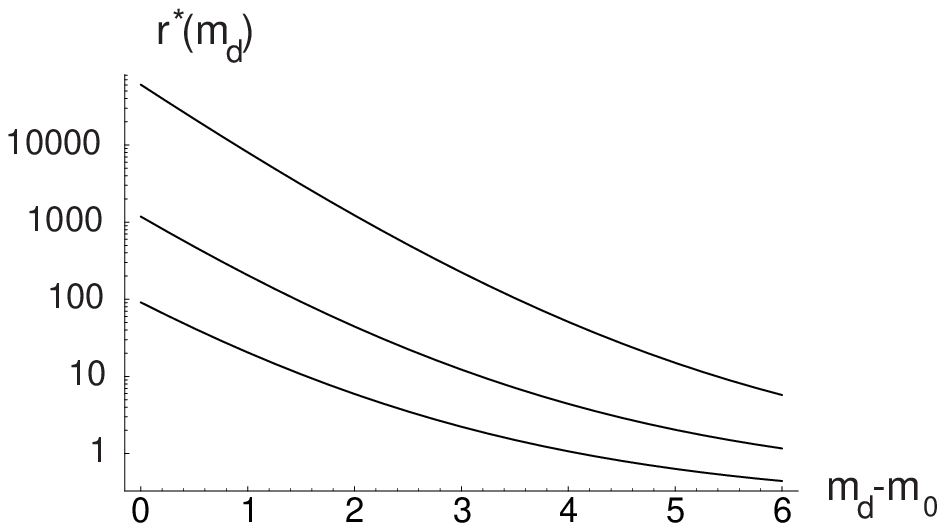}}
{\bf Fig.~7:} \small{Dependence of the cross-over value $r^*(m_d)$ 
separating the two power laws (\ref{74}) and (\ref{75}) for the statistics
of the number of observable events,
for $\gamma=1.25$ and $n=0.9;0.8;0.7$ (top to bottom). }
\end{quote}

\clearpage

\begin{quote}
\centerline{
\includegraphics[width=14cm]{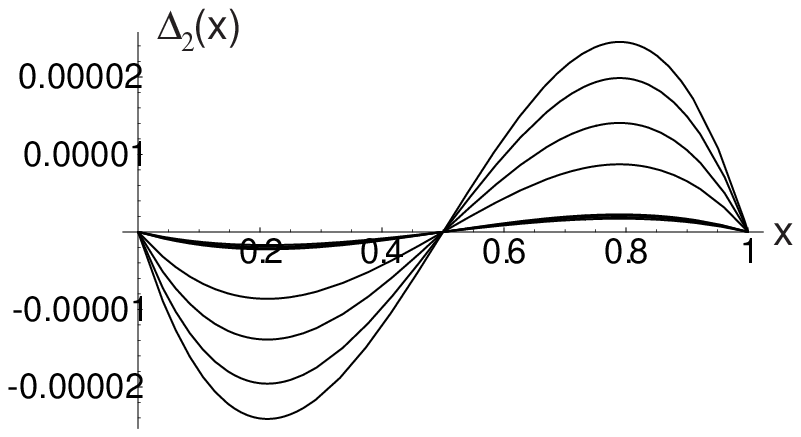}}
{\bf Fig.~8:} \small{Dependence of the difference $\Delta_2(x;m_d)$
defined in (\ref{85}) as a function of $x$
for the same parameters as in Figure~6,
demonstrating the high accuracy of the quadratic approximation (\ref{83}).}
\end{quote}

\clearpage

\begin{quote}
\centerline{
\includegraphics[width=14cm]{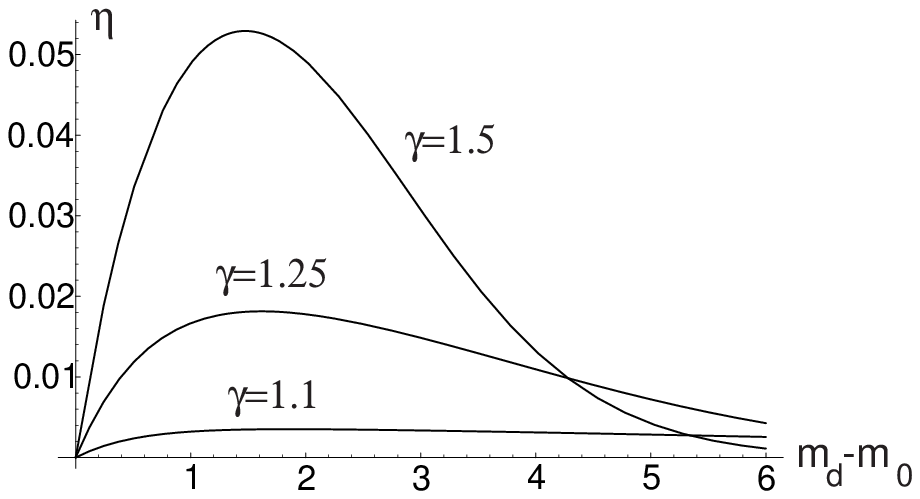}}
{\bf Fig.~9:} \small{Dependence of the 
shift $\eta(m_d)$ to the effective branching rate for
observable events as a function of $m_d-m_0$ for different 
values of $\gamma$ and for $n=0.9$.}
\end{quote}

\end{document}